# Nonreciprocal thermal transport in a multiferroic helimagnet

[1]Department of Basic Science, University of Tokyo, Tokyo 153-8902, Japan
[2]Institute for Materials Research, Tohoku University, Sendai 980-8577, Japan
[3]PRESTO, Japan Science and Technology Agency (JST), Kawaguchi 332-0012, Japan

Yuji Hirokane[1], Yoichi Nii[2,3], Hidetoshi Masuda[2], Yoshinori Onose[2]

**Breaking of spatial inversion symmetry (SIS) induces unique phenomena in condensed matter[1,2]. Besides the classic examples such as natural optical activity and piezoelectricity, the spin-orbit interaction further enriches the effect of SIS breaking, as exemplified by the Rashba effect[3]. In particular, by combining this symmetry with magnetic fields or another type of time-reversal symmetry (TRS) breaking, noncentrosymmetric materials can be made to exhibit nonreciprocal responses, which are responses that differ for rightward and leftward stimuli[4-6]. For example, resistivity becomes directionally dependent; that is to say, rectification appears in noncentrosymmetric materials in a magnetic field[7-10]. However, the effect of SIS breaking on thermal transport remains to be elucidated. Here we show nonreciprocal thermal transport in the multiferroic helimagnet TbMnO$_3$[11]. The longitudinal thermal conductivity depends on whether the thermal current is parallel or antiparallel to the vector product of the electric polarization and magnetization. This phenomenon is thermal rectification that is controllable with external fields in a uniform crystal. This discovery may pave the way to thermal diodes with controllability and scalability.**

Nonreciprocal responses in noncentrosymmetric materials have been investigated widely[6]. Nonreciprocal responses were first studied in terms of optical properties[4,5]. When time-reversal symmetry (TRS) and spatial inversion symmetry (SIS) are simultaneously broken, the optical properties such as absorption and luminescence become dependent on the sign of the optical wave vector irrespective of the polarization. Similar directional dependences were observed for magnon and phonon propagation, and electronic transport[7-10,12-15]. The symmetrical rule for the unidirectional response is common among these nonreciprocities. In chiral materials, in which any mirror symmetry is broken, directional dependence is observed along the magnetization **M** for any type of nonreciprocities. On the other hand, for polar materials, which have finite electric polarization **P**, the directional dependence is observed along **P** × **M**. In this paper, we study nonreciprocal thermal transport in a polar system.

We chose the multiferroic spiral magnet TbMnO$_3$ as a sample. TbMnO$_3$ shows transverse spiral magnetic order with the propagation vector along the *b*-axis, accompanied by **P** along the *c*-axis[15]. We measured nonreciprocal thermal transport in magnetic fields along the *a*-axis, in which a conical magnetic structure is realized (Fig. 1a). Let us discuss the phononic state in this magnetic structure to study the thermal transport microscopically. As Hamada and coworkers theoretically suggested, the breaking of SIS induces phonon angular momentum **L**, similarly to the Rashba and Dresselhaus effects in electronic states[16]. In polar systems, the induced **L** at momentum **k** is along **k** × **P**. Because **L** is caused by circular or ellipsoidal phonon polarization in the plane perpendicular to **L**, it originates from the mixing of two linearly polarized phonon modes in the plane. In the low-energy acoustic branches, these two modes respectively correspond to the longitudinal mode and one of the transverse modes with phonon polarization along **P**. The solid lines in Fig. 1b show the expected acoustic phonon modes for ferroelectric TbMnO$_3$ at zero field. The longitudinal mode and one of the transverse modes have the phonon angular momenta $L_0$ and $-L_0$, respectively, along **k** × **P** as a result of mixing. On the other hand, the other transverse mode has linear polarization along **k** × **P** and, therefore, the phonon angular momentum is zero. When **k** changes its sign, **L** is reversed, but the energy is unchanged. By the application of the magnetic field along the *a*-axis, the energies of phonon modes shift depending on **L** (dashed lines in Fig. 1b), owing to the so-called Raman-type spin-phonon interaction $H_{\text{Raman}} = K\,\mathbf{m}\cdot\mathbf{L}$[17,18], where **m** is a magnetic moment. Thus, the phonon energy dispersion becomes asymmetric in the magnetic field. This asymmetry of phononic dispersion is the origin of thermal nonreciprocity.

Thermal conductivity is usually measured in an experimental setup where one side of the sample is attached to a thermal bath and the other side to a heater (Fig. 1c). Because one needs to modify the experimental setup to reverse the heat current, it is difficult to estimate small rectification precisely. Here we demonstrate thermal nonreciprocity in the multiferroic spiral magnet TbMnO$_3$ by a comparison of the thermal conductivities for $\pm P$ and $\pm M$ states, without changing the direction of the thermal current. The thermal conductivity is expected to be rectified along the *b*-axis (||**P** × **M**). Reversal of thermal current is equivalent to **P** (**M**) reversals and 180-degree rotation of experimental setup around **M** (**P**). The 180-degree rotation does not change any physical responses and, therefore, the reversal of **P** or **M** corresponds to reversal of the thermal current. Thus, we can estimate a small thermal rectification by comparative measurements with positive and negative **P** and **M**. For this purpose, we attached a pair of electrodes to control the electric polarization in addition to the conventional thermal transport setup, as shown in Fig. 1c.

TbMnO$_3$ shows several magnetic phase transitions in the low temperature region. In Fig. 2a, we show the temperature dependence of magnetization at 0.5 T for TbMnO$_3$. The anomalies at 7 K, 27 K, and 43 K indicate the magnetic transitions[11,19]. The manganese magnetic moments order incommensurately below 43 K. While the magnetic moments are almost collinear above 27K, a cycloidal magnetic structure emerges, and ferroelectric polarization is induced along the *c*-axis below 27 K (Fig. 2b). The increase at 7 K corresponds to the incommensurate order of Tb moments. It should be noted that terbium oxides frequently show gigantic magnetoelastic phenomena[20,21]. The thermal nonreciprocity in the present study also seems to be related to the strong magnetoelastic coupling of terbium moments, as discussed later. The temperature dependence of thermal conductivity is shown in Fig. 2c. The thermal conductivity is relatively small and monotonically decreases with lowering temperature. A previous study suggests that these are caused by strong scattering due to terbium moments[22].

Next, let us discuss the magnetic field dependence of the magnetic and thermal properties at 4.2 K in order to demonstrate the thermal nonreciprocity. Figure 3a shows the magnetization as a function of the magnetic field along the *a*-axis at 4.2 K. The magnetization shows a step-like increase around 2 T. Because the c-axis polarization due to the spiral spin structure of manganese moments is robust in this magnetic field range[23], the transition is relevant to the ferromagnetic ordering of terbium moments. In Fig. 3b, we show the magnetic field dependence of thermal conductivity at 4.2 K. The thermal conductivity shows spike-like anomalies at the magnetic transition fields. A small peak is also observed at 0T. Small hysteresis is observed in the low-field region below the transition field. In Fig. 3c, we plot the magnetic field dependence of thermal conductivity above the metamagnetic transition field on a magnified scale for the $\pm P$ states. Importantly, it shows an asymmetric magnetic field dependence, and the asymmetry is reversed when the electric polarization is reversed. To scrutinize this phenomenon, we show in Fig. 3d the asymmetric part of the magnetic-field-dependent thermal conductivity: $\frac{\Delta \kappa}{\kappa} = (\kappa_+(H) - \kappa_-(-H))/(\kappa_+(H) + \kappa_-(-H))$ for the $\pm P$ states. Here, $\kappa_+(H)$ and $\kappa_-(H)$ are the thermal conductivity at a magnetic field $H$ measured in field-increasing and -decreasing runs, respectively. $\Delta \kappa / \kappa$ for +*P* shows an abrupt decrease around the magnetic transition and is almost constant in the high field region. When the polarization is reversed, $\Delta \kappa / \kappa$ changes its sign. These observations indicate that the thermal conductivity depends on whether the thermal current is parallel or anti-parallel to **P** × **M**. As mentioned above, various kinds of nonreciprocity were observed along **P** × **M** for samples with finite **P** and **M**. The present observation seems to correspond to the thermal version of nonreciprocal response. The thermal nonreciprocity is particularly enhanced when the terbium moments are fully polarized.

In Fig. 3d, we show the thermal current dependence of $\frac{\Delta\kappa_{av}}{\kappa}$ at 4.2 K for the $\pm P$ states. Here, $\Delta\kappa_{av}/\kappa$ is the average of $\Delta\kappa/\kappa$ in the region $3T \leq H \leq 5T$ and $-\Delta\kappa/\kappa$ in the region $-5T \leq H \leq -3T$. When the thermal current is too large, we cannot maintain the sample temperature. When it is too small, we cannot measure the thermal conductance accurately. At present, we measured thermal conductivity in the thermal-current range between $1.6\times10^2$ Wm$^{-2}$ and $3.2 \times10^2$ Wm$^{-2}$. The variation of $\Delta\kappa_{av}/\kappa$ is within the error bar in this range. Nevertheless, $\Delta\kappa$ is reversed by the reversal of **P**, which corresponds to reversal of the thermal current as mentioned above. Therefore, these observations indicate that the thermal conductivity depends on the sign of the thermal current but not on the magnitude.

The nonreciprocity of thermal conductivity $\frac{\Delta\kappa}{\kappa}$ at various temperatures is shown in Fig. 4a-h. A notable nonreciprocal response was also observed at 3.1 K. As the temperature is increased, the magnitude is decreased, and the step-like feature at the magnetic transition becomes obscure. The nonreciprocity almost vanished above 8.4 K. In Fig. 4i, we show the temperature dependence of $\Delta\kappa_{av}/\kappa$. It emerges around 8 K and increases with decreasing temperature. The onset temperature is close to the transition temperature of terbium. The magnetic field variation of thermal conductivity owing to the terbium moments is quite small above 8 K (See Fig. S1 in supplementary information). These characteristics also imply that the terbium moments are related to the origin of the nonreciprocal thermal transport.

Finally, let us discuss the origin of the nonreciprocal thermal transport. As mentioned above, the nonreciprocal signal correlates with the terbium magnetic state. Nevertheless, the terbium magnetic excitation hardly contributes to the thermal transport because it has a dispersionless local nature[24], and cannot carry heat. Therefore, inelastic scattering of phonons induced by the terbium moments seems to be the most plausible origin. When a phonon with angular momentum $L$ is scattered by a terbium moment with total angular momentum $J$, the sum of $L$ and $J$ must be conserved through the entire scattering process. Therefore, the transition matrix is expressed as:

$$< L + q, J - q|\mathrm{H}'|L, J >,$$

where H' is the interaction between the phonon and magnetic moment, and $q$ is the angular momentum transferred from the terbium magnetic moment to the phonon. As shown in Fig. 3d, the nonreciprocal thermal transport is notable when the terbium moment is ferromagnetically polarized. In this case, the initial state of $J$ is maximized. In other words, $J$ should be decreased and $L$ should be increased in the course of scattering. Therefore, the scattering by terbium moments is forbidden if the initial state is $L = +1$ or the final state is $L = -1$. As mentioned above, when SIS is broken, some phonon modes have finite angular momentum. Therefore, the effect of the selection rule should appear in the phononic properties.

To understand how the terbium magnetic scattering affects the thermal transport properties, let us calculate the nonreciprocal thermal transport for a simplified model. This is just for a qualitative understanding of nonreciprocal thermal transport in this system while a more elaborate theory is needed for the quantitative understanding. We assume one-dimensional asymmetric phonon bands as indicated by the dotted lines in Fig. 1 b and an induced phonon momentum magnitude $|L_0| = 1$. Therefore, the phonon angular momenta for three acoustic branches are $L = 0, \pm 1$. The phonon density for the angular momentum $L$ at momentum $k$ under a temperature gradient is expressed as $g(k, L) = g_0(k, L) + g'(k, L)$, where $g_0$ and $g'$ are the phonon density in thermal equilibrium and the deviation, respectively. Assuming the relaxation time approximation, we get $-\frac{g'}{\tau} = -v \frac{\partial g_0}{\partial T} \frac{\partial T}{\partial x}$. Here, $\tau, v,$ and $x$ are the relaxation time, phonon velocity, and position, respectively. When $k$ and $v$ are both positive (negative), $g'(k, L) > 0$ ($g'(k, L) < 0$) and the relaxation process is caused by the scattering whose initial state (final state) is $|k, L>$. Therefore, for $\frac{\partial T}{\partial x} > 0$ ($\frac{\partial T}{\partial x} < 0$), the terbium magnetic scattering is forbidden for the $L=1$ ($L=-1$) mode in the range $k > 0$, and for the $L = -1$ ($L=1$) mode in the range $k<0$. For simplicity, we assume linear dispersion relations in the phonon band structure. The velocities for the $L=+1$, 0, -1 modes are $v_{+,+}, v_0$, and $v_{-,+}$ for $k>0$ and $-v_{+,-}, -v_0$, and $-v_{-,-}$ for $k<0$. The scattering rate is assumed to be $\frac{1}{\tau} = \frac{1}{\tau_0} + \frac{1}{\tau_{mag}} \cong \frac{1}{\tau_0 - \tau_0^2/\tau_{mag}}$ when the terbium magnetic scattering is not forbidden, and $\frac{1}{\tau} = \frac{1}{\tau_0}$ when it is forbidden. Here, $\frac{1}{\tau_0}$ and $\frac{1}{\tau_{mag}}$ are the nonmagnetic and magnetic scattering rate, respectively. We assume $\frac{1}{\tau_0} \gg \frac{1}{\tau_{mag}}$ and that they are constant for simplicity. For

$\frac{\partial T}{\partial x} > 0$, the thermal conductivity can be expressed as

$$\kappa \cong \int_{k>0,k<0} \epsilon v_0^2 \left(\tau_0 - \frac{\tau_0^2}{\tau_{mag}}\right) \frac{\partial g_0}{\partial T} dk + \int_{k>0} \epsilon v_{-,+}^2 \left(\tau_0 - \frac{\tau_0^2}{\tau_{mag}}\right) \frac{\partial g_0}{\partial T} dk + \int_{k>0} \epsilon v_{+,+}^2 \tau_0 \frac{\partial g_0}{\partial T} dk$$

$$+ \int_{k<0} \epsilon v_{+,-}^2 \left(\tau_0 - \frac{\tau_0^2}{\tau_{mag}}\right) \frac{\partial g_0}{\partial T} dk + \int_{k<0} \epsilon v_{-,-}^2 \tau_0 \frac{\partial g_0}{\partial T} dk.$$

Similarly, for $\frac{\partial T}{\partial x} < 0$, we get

$$\kappa \cong \int_{k>0,k<0} \epsilon v_0^2 \left(\tau_0 - \frac{\tau_0^2}{\tau_{mag}}\right) \frac{\partial g_0}{\partial T} dk + \int_{k>0} \epsilon v_{-,+}^2 \tau_0 \frac{\partial g_0}{\partial T} dk + \int_{k>0} \epsilon v_{+,+}^2 \left(\tau_0 - \frac{\tau_0^2}{\tau_{mag}}\right) \frac{\partial g_0}{\partial T} dk$$

$$+ \int_{k<0} \epsilon v_{+,-}^2 \tau_0 \frac{\partial g_0}{\partial T} dk + \int_{k<0} \epsilon v_{-,-}^2 \left(\tau_0 - \frac{\tau_0^2}{\tau_{mag}}\right) \frac{\partial g_0}{\partial T} dk.$$

Then, the nonreciprocity is

$$\Delta\kappa = \frac{\tau_0^2}{\tau_{mag}} \left( \int_{k>0} \epsilon v_{+,+}^2 \frac{\partial g_0}{\partial T} dk + \int_{k<0} \epsilon v_{-,-}^2 \frac{\partial g_0}{\partial T} dk - \int_{k<0} \epsilon v_{+,-}^2 \frac{\partial g_0}{\partial T} dk - \int_{k>0} \epsilon v_{-,+}^2 \frac{\partial g_0}{\partial T} dk \right).$$

This relation is finite when $v_{+,+} \neq v_{+,-}$ or $v_{-,+} \neq v_{-,-}$. Thus, the scattering by ferromagnetic terbium moments and simultaneous breaking of SIS and TRS induce the nonreciprocal thermal transport. This is independent of the magnitude of the thermal current density. If the magnitude of the induced phonon angular momentum $L_0$ is less than 1, the nonreciprocity is reduced but still finite unless $L_0$ becomes zero. This simple model is consistent with the experimental observations; the nonreciprocity is fairly enhanced by the ferromagnetic alignment of terbium moments and $\Delta\kappa/\kappa$ is independent of thermal current density.

In summary, we have demonstrated nonreciprocal thermal transport in multiferroic TbMnO$_3$; the thermal transport is rectified along the vector product of the polarization and magnetization. The thermal rectification seems useful for the efficient management of heat in various circumstances[25]. Similar thermal rectifications were previously realized by using combinations of two different materials, or asymmetric shapes of media[25-30]. In these cases, the forward and reverse directions are determined by structural asymmetries, and the directionality cannot be reversed by external fields. In addition, an asymmetric structure limits the scale of the device. In the present case, the thermal rectification is realized in a uniform crystal and can be reversed by the external electric or magnetic fields. If the magnitude of thermal nonreciprocity becomes large enough to make practical applications feasible, the scalability of the uniform crystal and the ability to control the phenomenon with the external fields are significant advantages over the previous approaches. Thus, the present results suggest that the exploration of thermal properties in symmetry-broken materials is a new research strategy that should be investigated toward the realization of thermal diode devices.

## Methods

A TbMnO$_3$ single crystal was grown by means of the floating zone technique. The conditions were almost the same as those in previous reports[15,23]. Magnetic susceptibility was measured with the use of a superconducting quantum interference device (SQUID). Thermal conductivity measurements were performed by means of the steady-state method in a superconducting magnet. The measurement atmosphere was evacuated down to 3×10$^{-3}$Pa. Heat current was generated by chip resistance. The thermal gradient was measured by two thermometers (CERNOX, Lake Shore Cryotronics, Inc.). To control the electric polarization, a pair of Au/Ti electrodes were formed on the surface of the thermal transport sample. The thicknesses of the Au and Ti were 20 nm and 2 nm, respectively.

## Data availability

Data that support the findings of this study are available from the corresponding author upon reasonable request.

## Acknowledgements

The authors thank S. Murakami for fruitful discussions. This work was supported in part by JSPS KAKENHI (grant numbers JP16H04008, JP17H05176, JP18K13494), PRESTO (grant number JPMJPR19L6), the Murata Science Foundation, and the Mitsubishi foundation.

## Author contributions

Y. H. carried out the crystal growth, magnetization measurement, and thermal transport measurement with assistance from Y.N. and H.M. Y.O. conceived and supervised the project. Y.O. wrote the paper through the discussion and assistance from Y. H. Y. N., and H. M.

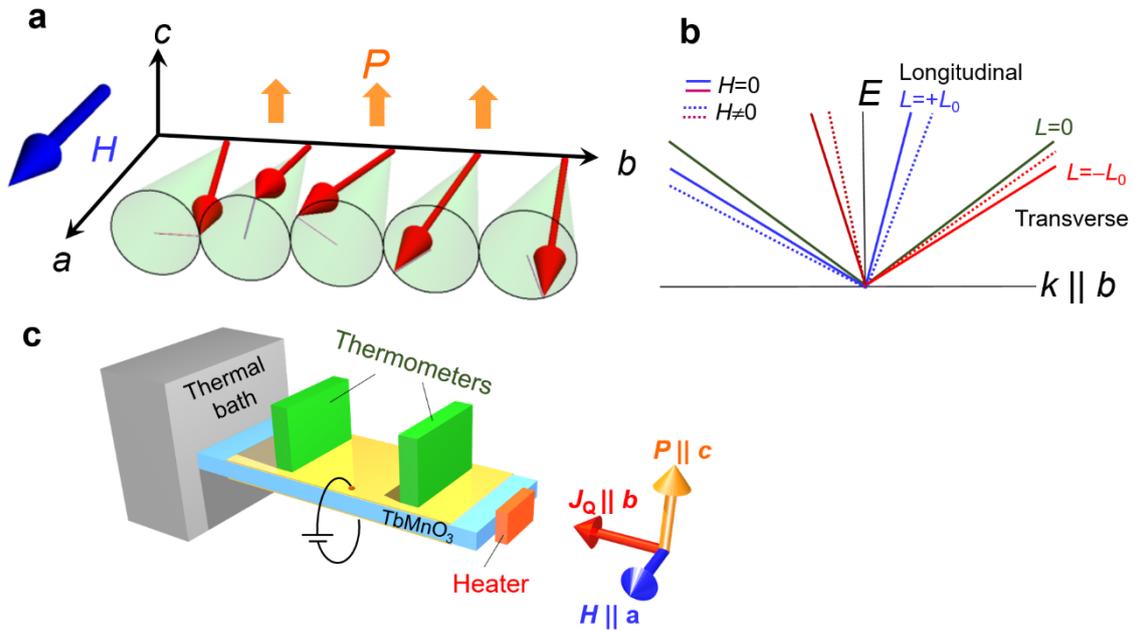

Figure 1 Magnetic and phononic states of multiferroic TbMnO$_3$ and experimental setup for nonreciprocal thermal transport. **a,** Illustration of magnetic structure in a magnetic field $H$ along the $a$-axis. **b,** Expected acoustic phonon modes at $H=0$ (solid lines) and finite $H$ along the $a$-axis (dotted lines). **c,** Illustration of the experimental setup for nonreciprocal thermal transport measurement.

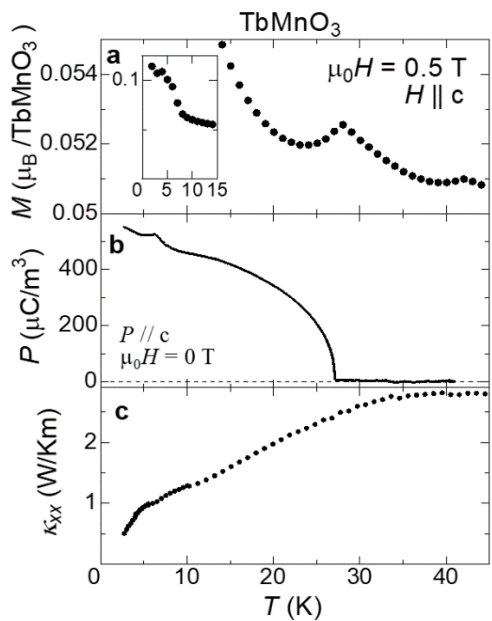

Figure 2 Temperature dependences of magnetic, electric, and thermal transport properties of TbMnO$_3$ **a,** Temperature dependence of magnetization at $H=0.5$ T along the $c$-axis. **b,** Temperature dependence of electric polarization along the c-axis

at $H=0$. **c,** Temperature dependence of thermal conductivity at $H=0$.

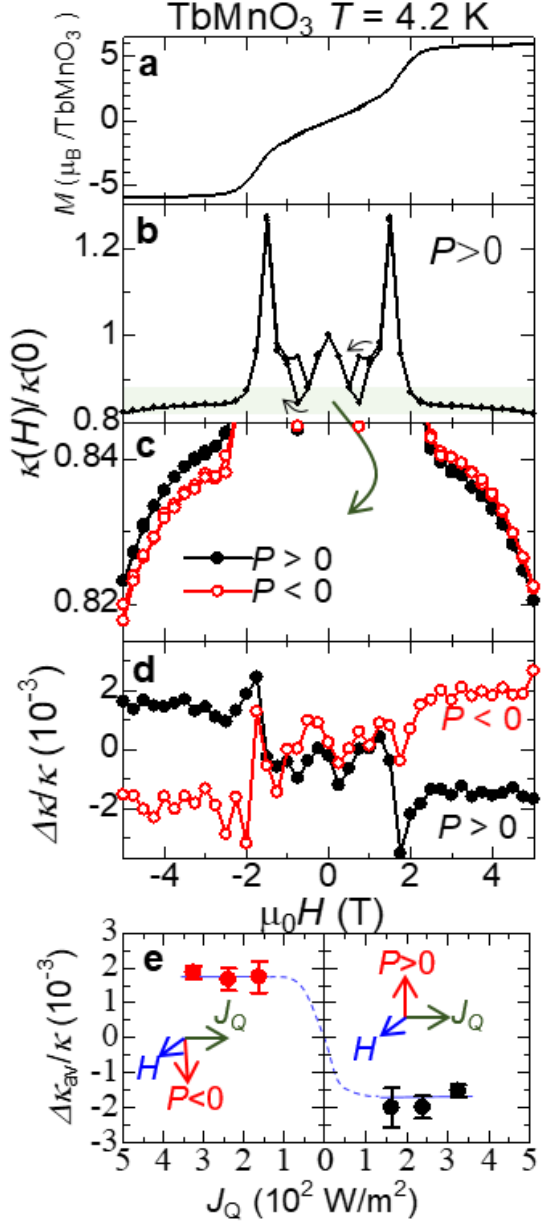

Figure 3 Nonreciprocal thermal transport at 4.2 K. **a,** Magnetization curve along the $a$-axis at 4.2 K. **b,c,** Magnetic field variation of thermal conductivity at 4.2 K for $H \parallel a$. In **c,** the thermal conductivity scale is magnified around $\kappa(H)/\kappa(0) = 0.83$. **d,** Magnetic field dependence of thermal nonreciprocity $\frac{\Delta\kappa}{\kappa}$ for $\pm P$ and $H \parallel a$ at 4.2 K. **e,** Averaged thermal nonreciprocity $\Delta\kappa_{av}/\kappa$ as a function of thermal current density at 4.2 K. The error bars are estimated as the error of thermal conductivity from the temperature

fluctuation during the measurement divided by the square of number of averaged data points. Blue solid and dashed lines are merely guides for the eyes.

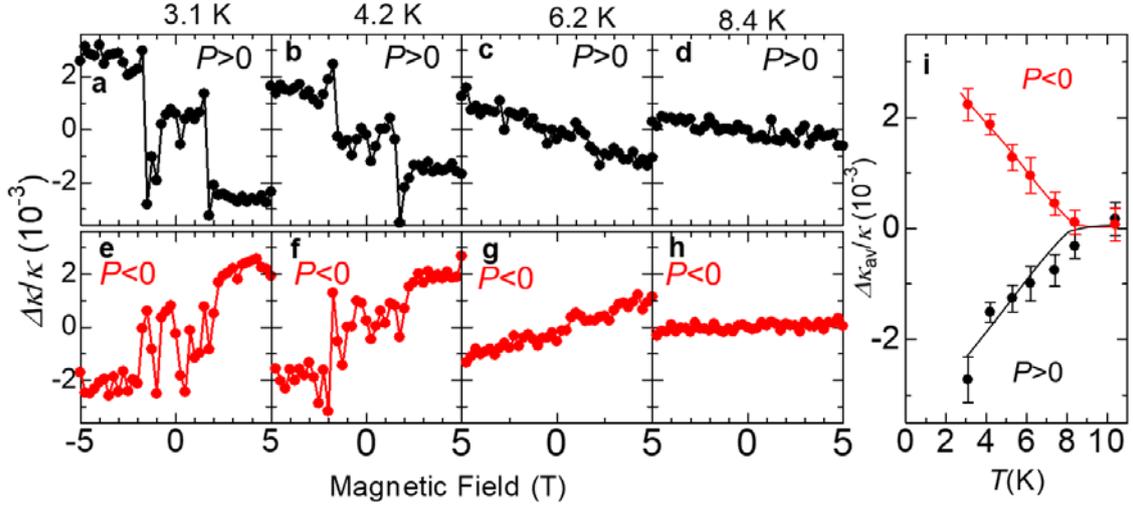

Figure 4 Temperature dependence of nonreciprocal thermal transport. **a-h**, Magnetic field variation of thermal nonreciprocity $\Delta\kappa/\kappa$ for $\pm P$ and $H \parallel a$ at various temperatures. **i,** Temperature dependence of averaged thermal nonreciprocity $\Delta\kappa_{av}/\kappa$ for $\pm P$. The error bars are estimated as the error of thermal conductivity from the temperature fluctuation during the measurement divided by the square of number of averaged data points. Solid lines are merely guides for the eyes

# Supplementary Information for
# "Nonreciprocal thermal transport in a multiferroic helimagnet"

[1]Department of Basic Science, University of Tokyo, Tokyo 153-8902, Japan
[2]Institute for Materials Research, Tohoku University, Sendai 980-8577, Japan

Yuji Hirokane[1], Yoichi Nii[2], Hidetoshi Masuda[2], Yoshinori Onose[2]

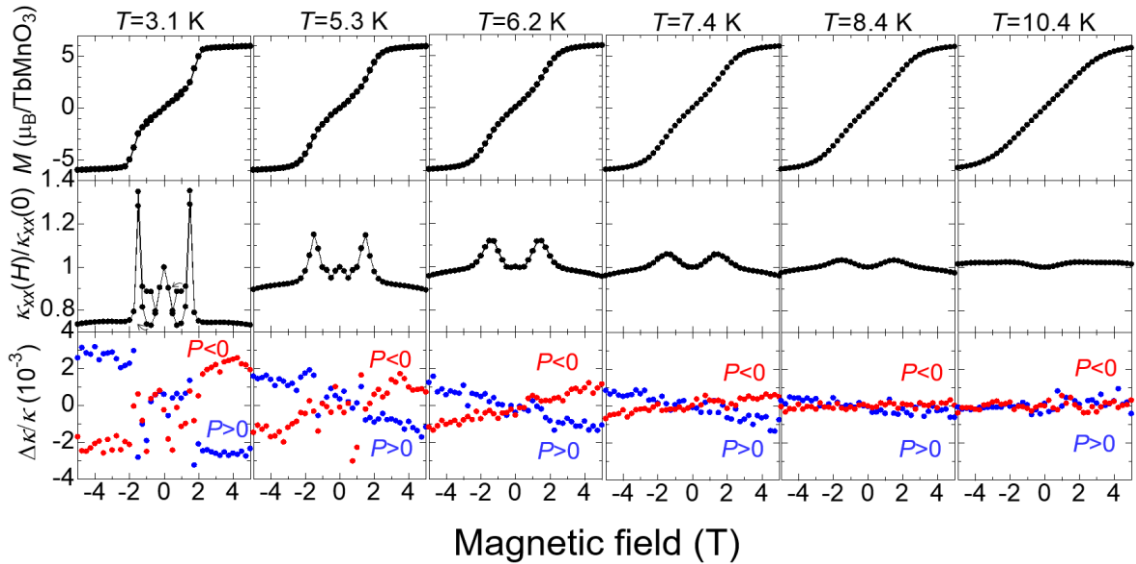

Figure S1: Magnetic and thermal transport properties at various temperatures. Magnetic field variations of magnetization, thermal conductivity, and thermal nonreciprocity for $\pm P$ at $T$=3.1K, 5.3K, 6.2 K, 7.4 K, 8.4 K, and 10.4 K.